\documentclass[twocolumn,showpacs,amsmath,amssymb,prl]{revtex4}

\usepackage{graphicx}
\usepackage{dcolumn}
\usepackage{bm}

\begin{document}

\title{Test of Lorentz Invariance in Electrodynamics Using Rotating Cryogenic Sapphire Microwave Oscillators}
\author{Paul L. Stanwix$^1$}
\email{pstanwix@physics.uwa.edu.au}
\author{Michael E. Tobar$^1$}
\email{mike@physics.uwa.edu.au}
\author{Peter Wolf$^{2,3}$}
\author{Mohamad Susli$^1$}
\author{Clayton R. Locke$^1$}
\author{Eugene N. Ivanov$^1$}
\author{John Winterflood$^1$}
\author{Frank van Kann$^1$}
\affiliation{\\ $^1$University of Western Australia, School of
Physics M013, 35 Stirling Hwy., Crawley 6009 WA, Australia\\
$^2$SYRTE, Observatoire de Paris, 61 Av. de l'Observatoire, 75014 Paris, France\\
$^3$Bureau International des Poids et Mesures, Pavillon de Breteuil,
92312 S\`evres Cedex, France}

\date{\today}

\begin{abstract}
We present the first results from a rotating Michelson-Morley
experiment that uses two orthogonally orientated cryogenic sapphire 
resonator-oscillators operating in whispering gallery modes near 10 GHz.
The experiment is used to test for violations of Lorentz Invariance
in the frame-work of the photon sector of the Standard Model
Extension (SME), as well as the isotropy term of the
Robertson-Mansouri-Sexl (RMS) framework. In the SME we set a new
bound on the previously unmeasured $\tilde{\kappa}_{e-}^{ZZ}$
component of $2.1(5.7)\times10^{-14}$, and set more stringent bounds
by up to a factor of 7 on seven other components. In the RMS a more
stringent bound of $-0.9(2.0)\times 10^{-10}$ on the isotropy
parameter, $P_{MM}=\delta -  \beta + \frac{1}{2}$ is set, which is more than a
factor of 7 improvement.
\end{abstract}

\maketitle

The Einstein Equivalence Principle (EEP) is a founding principle of relativity \cite{Will}. 
One of the constituent elements of EEP is Local Lorentz Invariance (LLI), which postulates
that the outcome of a local experiment is independent of the
velocity and orientation of the apparatus. The central importance of
this postulate has motivated tremendous work to
experimentally test LLI. Also, a number of unification theories
suggest a violation of LLI at some level. However, to
test for violations it is necessary to have an alternative theory to
allow interpretation of experiments \cite{Will}, and many have
been developed \cite{Robertson,MaS,LightLee,Ni,Kosto1,KM}. The
kinematical Roberson-Mansouri-Sexl (RMS) \cite{Robertson, MaS} framework postulates a
simple parameterization of the Lorentz transformations with
experiments setting limits on the deviation of those parameters from
their values in special relativity (SR). Because of their simplicity
they have been widely used to interpret
many experiments \cite{Brillet,Wolf,Muller,WolfGRG}. More recently,
a general Lorentz violating extension of the standard model of
particle physics (SME) has been developed \cite{Kosto1} whose
Lagrangian includes all parameterized Lorentz violating terms that
can be formed from known fields.

This work presents first results of a rotating lab experiment using cryogenic microwave oscillators. Previous non-rotating
experiments \cite{Lipa,Muller,Wolf04} relied on the earth's rotation
to modulate a Lorentz violating effect. This is not optimal for two
reasons. Firstly, the sensitivity is
proportional to the noise of the oscillators at the
modulation frequency, typically best for periods between 10 and
100 seconds. Secondly, the sensitivity is
proportional to the square root of the number of periods of the modulation signal, therefore taking a relatively long time to acquire sufficient data. Thus, by rotating the experiment the data integration rate is increased and the relevant signals are translated to the optimal operating regime \cite{Mike}.

Our experiment consists of two cylindrical sapphire resonators of 3 cm diameter and height supported by spindles within superconducting niobium cavities \cite{Giles}, and are oriented with their cylindrical axes orthogonal to each other in the horizontal plane.
Whispering gallery modes \cite{wgmode} are excited near 10 GHz, with a difference frequency of 226 kHz. The frequencies are stabilized using Pound locking, and amplitude variations
are suppressed using an additional control circuit. A detailed
description of such oscillators can be found in
\cite{Mann, Hartnett}. The resonators are
mounted in a common copper block, which provides
common mode rejection of temperature fluctuations. The structure is in turn mounted inside two successive stainless steel vacuum cylinders from a copper post, which provides the thermal connection between the cavities and the liquid helium bath. A foil heater and carbon-glass temperature sensor attached to the copper post controls the temperature set point
to 6 K with mK stability.

\begin{figure}
\begin{center}
\includegraphics[width=3in]{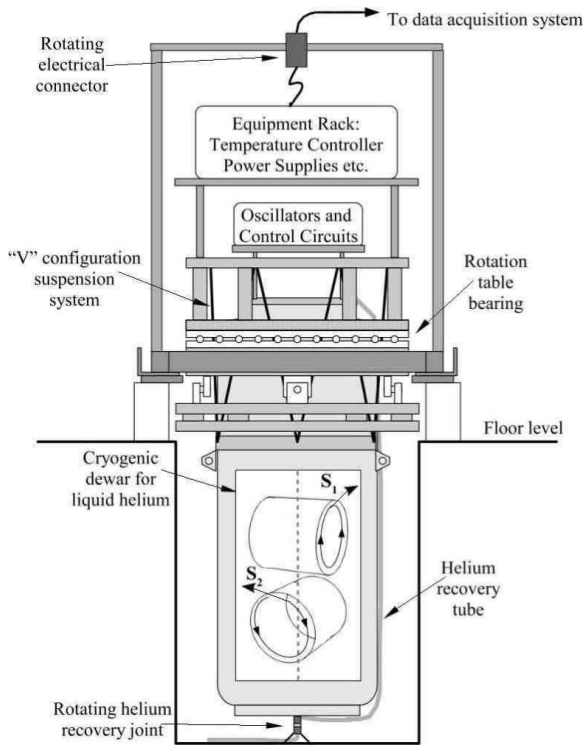}
\caption{Schematic of the cryogenic dewar, mounted in the rotation
table. Inside the dewar a schematic of the two orthogonally
orientated resonators is shown, along with the Poynting vectors of
propagation $S_{1}$ and $S_{2}$.} \label{fig:rotPic}
\end{center}
\end{figure}
\begin{figure}
\begin{center}
\includegraphics[width=3in]{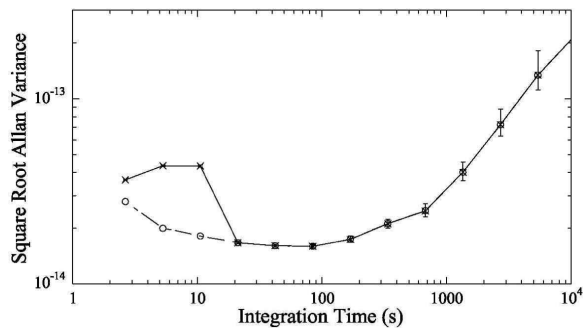}
\caption{Square Root Allan Variance fractional frequency instability measurement of the difference frequency when rotating (crosses) and stationary (circles). The hump at short integration times is due to systematic effects associated with the rotation of the experiment, with a period of 18 seconds. Above 18 seconds the instability is the same as when the experiment is stationary.}
\label{fig:stab}
\end{center}
\end{figure}
A schematic of the rotation system is shown in Fig.\ref{fig:rotPic}.
The cryogenic dewar along with the room
temperature oscillator and control electronics is
suspended within a ring bearing. A multiple "V" shaped suspension
made from elastic cord avoids high Q-factor pendulum
modes by ensuring that the cord has to stretch and shrink (providing
damping) for horizontal and vertical motion. The rotation system is driven by a microprocessor controlled stepper motor. A commercial 18 conductor slip ring connector, with a hollow through bore, transfers power and various signals to and from the rotating experiment. A mercury based rotating coaxial connector
transmits the difference frequency to a stationary frequency counter
referenced to an Oscilloquartz oscillator. The data acquisition
system logs the difference frequency as a function of orientation,
as well as monitoring systematic effects including the temperature
of the resonators, liquid helium bath level, ambient room
temperature, oscillator control signals, tilt, and helium return
line pressure.

Inside the sapphire crystals standing waves are set up with the dominant electric and
magnetic fields in the axial and radial directions respectively,
corresponding to a Poynting vector around the
circumference. The experimental observable is the difference
frequency, and to test for Lorentz violations the perturbation of
the observable with respect to an alternative test theory
must be derived. For example, in the photon sector of the SME
this may be calculated to first order as the
integral over the non-perturbed fields (Eq. (34) of \cite{KM}), and
expressed in terms of 19 independent variables
(discussed in more detail later). The change in orientation of the
fields due to the lab rotation and Earth's orbital and
sidereal motion induces a time varying modulation of the difference
frequency, which is searched for in the experiment. Alternatively,
with respect to the RMS framework, we analyze the change in resonator frequency as a function of the Poynting vector direction with respect to the velocity of the lab through the cosmic microwave background (CMB). The RMS parameterizes a possible Lorentz violation by a deviation of the parameters ($\alpha, \beta, \delta$) from their SR values ($-\frac{1}{2}, \frac{1}{2}, 0$). Thus, a complete verification of LLI in the RMS framework \cite{Robertson,MaS} requires a test of (i) the
isotropy of the speed of light ($P_{MM}=\delta -  \beta +
\frac{1}{2}$), a Michelson-Morley (MM) experiment \cite{MM}, (ii)
the boost dependence of the speed of light ($P_{KT}=\beta - \alpha -
1$), a Kennedy-Thorndike (KT) experiment \cite{KT} and (iii) the
time dilation parameter ($P_{IS}=\alpha + \frac{1}{2}$), an
Ives-Stillwell (IS) experiment \cite{IS,Saat}. Because our
experiment compares two cavities it is only sensitive to $P_{MM}$.

Fig.\ref{fig:stab} shows typical fractional frequency instability of the 226 kHz difference with respect to 10 GHz, and compares the instability when rotating and stationary. A minimum of $1.6\times10^{-14}$ is recorded at 40s. Rotation induced systematic effects degrade the stability up to 18s due to signals at the rotation frequency of $0.056 Hz$ and its harmonics. We have determined that tilt variations dominate the systematic by measuring the magnitude of the fractional frequency dependence on tilt and the variation in tilt at twice the rotation frequency, $2\omega_R (0.11 Hz)$, as the experiment rotates. We minimize the effect of tilt by manually setting the rotation bearing until our tilt sensor reads a minimum at $2\omega_R$. The latter data sets were up to an order of magnitude reduced in amplitude as we became more experienced at this process. The remaining systematic signal is due to the residual tilt variations, which could be further annulled with an automatic tilt control system. It is still possible to be sensitive to Lorentz violations in the presence of these
systematics by measuring the sidereal, $\omega_\oplus$ and semi-sidereal, $2\omega_\oplus$ sidebands about $2\omega_R$, as was done in \cite{Brillet}. The
amplitude and phase of a Lorentz violating signal is determined by fitting the parameters of Eq.\ref{nuTest} to the data, with the phase of the fit adjusted according to the test theory used.
\begin{equation}
\frac{\Delta\nu_0}{\nu_0} = A + B t + \sum_i C_i {\rm
cos}(\omega_{i}t + \varphi_i) + S_i {\rm sin}(\omega_{i}t +
\varphi_i) \label{nuTest}
\end{equation}
Here $\nu_0$ is the average unperturbed frequency of the two
sapphire resonators, and  $\Delta\nu_0$ is the perturbation of the
226 kHz difference frequency, $A$ and $B$ determine the frequency offset
and drift, and $C_i$ and $S_i$ are the amplitudes of a cosine and
sine at frequency $\omega_i$  respectively. In the final analysis we
fit 5 frequencies to the data, $\omega_i = (2\omega_R,
2\omega_R\pm\omega_\oplus, 2\omega_R\pm 2\omega_\oplus)$, as well as
the frequency offset and drift. The correlation coefficients between
the fitted parameters are all between $10^{-2}$ to $10^{-5}$. 
Since the residuals exhibit a significantly non-white behavior, 
the optimal regression method is weighted least squares (WLS) \cite{Wolf04}. 
WLS involves pre-multiplying both the experimental data and the model matrix by a
whitening matrix determined by the noise type of the residuals of an ordinary least squares analysis.

\begin{figure}
\begin{center}
\includegraphics[width=3in]{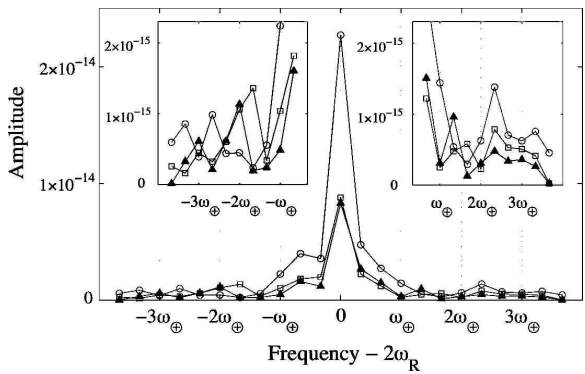}
\caption{Spectrum of amplitudes $\sqrt{C_i^2+S_i^2}$ calculated using WLS, showing systematic leakage about $2\omega_R$
for 2 data sets, data set 1 (3.6 days, circles), data set 5 (6.1
days, squares) and the combined data (18 days spanning 3 months, solid triangles). Here $\omega_\oplus$ is the sidereal frequency $(11.6 \mu Hz)$. By comparing a variety of data sets we have seen that leakage is reduced in longer data sets with lower systematics.
The insets show the typical amplitude away from the systematic, which have statistical uncetainties of order $10^{-16}$.}
\label{fig:data}
\end{center}
\end{figure}

We have acquired 5 sets of data over a period of 3 months beginning
December 2004, totaling 18 days.  The length of the sets (in days)
and size of the systematic are ($3.6, 2.3\times10^{-14}$), ($2.4,
2.1\times10^{-14}$), ($1.9, 2.6\times10^{-14}$), ($4.7, 1.4 \times
10^{-15}$), and ($6.1, 8.8 \times 10^{-15}$) respectively. We have
observed leakage of the systematic into the neighboring side bands
due to aliasing when the data set is not long enough or the
systematic is too large. Fig.\ref{fig:data} shows the total
amplitude resulting from a WLS fit to 2 of the data sets over a
range of frequencies about $2\omega_R$. It is evident that the
systematic of data set 1 at $2\omega_R$ is affecting the fitted
amplitude of the sidereal sidebands $2\omega_R\pm\omega_\oplus$ due
to its relatively short length and large systematics. By analyzing
all five data sets simultaneously using WLS the effective length of
the data is increased, reducing the width of the systematic
sufficiently as to not contribute significantly to the sidereal and semi-sidereal
sidebands.
\begin{table*}
\caption{\label{Tab1} Coefficients $C_i$ and $S_i$ in (1) for the
five frequencies of interest and their relation to the components of
the SME parameters $\tilde{\kappa}_{e-}$ and $\tilde{\kappa}_{o+}$,
derived using a short data set approximation including terms up to
first order in orbital velocity, where $\Phi_0$ is the phase of the
orbit since the vernal equinox (see \cite{SR2005} for details of the calculation). Note that for short data sets the upper and lower sidereal sidebands are redundant, which reduces the
number of independent measurements to 5. To lift the redundancy,
more than a year of data is required so annual offsets may be
de-correlated from the twice rotational and sidereal sidebands
listed.}
\begin{ruledtabular}
\begin{tabular}{ccccc}
$\omega_i$ & $C_i$ & $S_i$ \\
\hline
$2\omega_R$ & $0.21\tilde{\kappa}_{e-}^{ZZ}$ & - \\
$2\omega_R + \omega_\oplus$ & $2.5\times 10^{-5}\sin\Phi_0\tilde{\kappa}_{o+}^{XY}-1.0\times 10^{-5}\cos\Phi_0\tilde{\kappa}_{o+}^{YZ}$ & $-\cos\Phi_0\left[2.3\times 10^{-5}\tilde{\kappa}_{o+}^{XY}-1.0\times 10^{-5}\tilde{\kappa}_{o+}^{XZ}\right]$ \\
&$-0.27 \ \tilde{\kappa}_{e-}^{XZ}$ &$-0.27\tilde{\kappa}_{e-}^{YZ}$\\
$2\omega_R +2\omega_\oplus$ & $-2.1\times 10^{-5}\cos\Phi_0\tilde{\kappa}_{o+}^{XZ}$ & $-2.3\times 10^{-5}\sin\Phi_0\tilde{\kappa}_{o+}^{XZ}$ \\
&$+2.3\times 10^{-5}\sin\Phi_0\tilde{\kappa}_{o+}^{YZ}-0.11(\tilde{\kappa}_{e-}^{XX}-\tilde{\kappa}_{e-}^{YY})$ &$-2.1\times 10^{-5}\cos\Phi_0\tilde{\kappa}_{o+}^{YZ}-0.23 \ \tilde{\kappa}_{e-}^{XY}$\\
$2\omega_R - \omega_\oplus$ & $-0.31C_{2\omega_R +\omega_\oplus}$ & $0.31S_{2\omega_R +\omega_\oplus}$ \\
$2\omega_R -2\omega_\oplus$ & $9.4\times 10^{-2}C_{2\omega_R +2\omega_\oplus}$ & $-9.4\times10^{-2}S_{2\omega_R +2\omega_\oplus}$ \\
\end{tabular}
\end{ruledtabular}
\end{table*}

In the photon sector of the SME \cite{KM}, Lorentz violating terms
are parameterized by 19 independent components, which are in general
grouped into three traceless and symmetric $3\times 3$ matrices
($\tilde{\kappa}_{e+}$, $\tilde{\kappa}_{o-}$, and
$\tilde{\kappa}_{e-}$), one antisymmetric
matrix($\tilde{\kappa}_{o+}$) and one additional scalar, which all
vanish when LLI is satisfied. To derive the expected signal in the SME we use the method of \cite{KM, WolfGRG} to calculate the frequency of each resonator in the SME and in the resonator frame. We then transform to the standard celestial frame used in the SME \cite{KM} taking into account the  rotation in the laboratory frame in a similar way to \cite{TobarPRD}. 
The resulting relation between the
parameters of the SME and the $C_i$ and $S_i$ coefficients are given
in Tab.\ref{Tab1} which, for short data sets, were calculated using
the leading order expansion at the annual phase position of the
data. The 10 independent components of
$\tilde{\kappa}_{e+}$ and $\tilde{\kappa}_{o-}$ have been
constrained by astronomical measurements to $< 2\times 10^{-32}$
\cite{KM,Kost01}. Seven components of $\tilde{\kappa}_{e-}$ and
$\tilde{\kappa}_{o+}$ have been constrained in optical and microwave
cavity experiments \cite{Muller,Wolf04} at the $10^{-15}$ and
$10^{-11}$ level respectively, while the scalar
$\tilde{\kappa}_{tr}$ component recently had an upper limit set of
$< 10^{-4}$ \cite{TobarPRD}. The remaining
$\tilde{\kappa}_{e-}^{ZZ}$ component could not be previously
constrained in non-rotating experiments \cite{Muller,Wolf04}.

In contrast, our rotating experiment is sensitive to $\tilde{\kappa}_{e-}^{ZZ}$. However, it appears only at $2\omega_R$, which is dominated by systematic effects. From our combined analysis of all data sets, and using the relation to $\tilde{\kappa}_{e-}^{ZZ}$ given in Tab.\ref{Tab1}, we determine a value for $\tilde{\kappa}_{e-}^{ZZ}$
of $4.1(0.5)\times10^{-15}$. However, since we do not know if the
systematic has canceled a Lorentz violating signal at $2\omega_R$, we cannot reasonably claim this as an upper limit. Since we have five individual data sets, a limit can be set  by treating the $C_{2\omega_R}$ coefficient as a statistic. The phase of the systematic depends on the initial experimental conditions, and is random across the data sets. Thus, we have five values of $C_{2\omega_R}$, ($\{-4.2,11.4, 21.4, 1.3, -8.1\}$ in $10^{-15}$). If we take the mean of these coefficients, the systematic signal will cancel if its phase is random, but the possible Lorentz violating signal (with constant phase) will not. Thus a limit can be set by taking the mean and standard deviation of the five coefficient of $C_{2\omega_R}$. This gives a more conservative bound of $2.1(5.7)\times 10^{-14}$, which includes zero. Our experiment is also sensitive to all other seven components of $\tilde{\kappa}_{e-}$ and $\tilde{\kappa}_{o+}$ (see Tab.\ref{Tab1}) and improves present limits by up to a factor of 7, as shown in Tab.\ref{Tab2}.

\begin{table}
\caption{\label{Tab2}Results for the SME Lorentz violation
parameters, assuming no cancelation between the isotropy terms
$\tilde{\kappa}_{e-}$ (in $10^{-15}$) and first order boost terms
$\tilde{\kappa}_{o+}$ (in $10^{-11}$) \cite{Lipa}.}
\begin{ruledtabular}
\begin{tabular}{ccccc}
& $\tilde{\kappa}_{e-}^{XY}$ & $\tilde{\kappa}_{e-}^{XZ}$ & $\tilde{\kappa}_{e-}^{YZ}$ & $(\tilde{\kappa}_{e-}^{XX}-\tilde{\kappa}_{e-}^{YY})$ \\
\hline
this work & -0.63(0.43) & 0.19(0.37) & -0.45(0.37) & -1.3(0.9)\\
from \cite{Wolf04} & -5.7(2.3) & -3.2(1.3) & -0.5(1.3) & -3.2(4.6)\\
\hline \hline
&  $\tilde{\kappa}_{e-}^{ZZ}$ & $\tilde{\kappa}_{o+}^{XY}$ & $\tilde{\kappa}_{o+}^{XZ}$ & $\tilde{\kappa}_{o+}^{YZ}$ \\
\hline
this work& $21(57)$ & 0.20(0.21) & -0.91(0.46) & 0.44(0.46) \\
from \cite{Wolf04}  & $-$ & -1.8(1.5) & -1.4(2.3) & 2.7(2.2)
\end{tabular}
\end{ruledtabular}
\end{table}

In the RMS frame-work, a frequency shift due to a putative Lorentz
violation is given by Eq.\ref{RMS} \cite{Wolf,WolfGRG},
\begin{equation}
\frac{\Delta\nu_0}{\nu_0} = \frac{P_{MM}}{2\pi
c^2}\left[\oint{\left(\bf{v}.\hat{\bf{\theta}}_{1}\right)^2
d\varphi_1}-\oint{\left(\bf{v}.\hat{\bf{\theta}}_{2}\right)^2
d\varphi_2}\right] \label{RMS}
\end{equation}
Where $\bf{v}$ is the velocity of the preferred frame wrt the CMB,
$\hat{\bf{\theta}}_{j}$ is the unit vector in the direction of the
azimuthal angle (direction of propagation) of each resonator
(labeled by subscripts 1 and 2), and $\varphi$ is the azimuthal
variable of integration in the cylindrical coordinates of each
resonator. Perturbations due to Lorentz violations occur at the same
five frequencies as the SME, but for the RMS analysis we do not
consider the $2\omega_R$ frequency due to the large systematic, as
we only need to put a limit on one parameter. The dominant
coefficients are due to only the cosine terms with respect to the
CMB right ascension, $Cu_i$, which are shown in Tab.\ref{Tab3}.

\begin{table}
\caption{\label{Tab3} Dominant coefficients in the RMS, using a
short data set approximation calculated from Eq.\ref{RMS}. The measured values of $P_{MM}$ (in $10^{-10}$) are shown together with
the statistical uncertainties in the bracket. From this data the
measured and statistical uncertainty of $P_{MM}$ is determined to be
$-0.9(2.0)\times10^{-10}$, which represents more than a factor of 7
improvement over previous results $2.2(1.5)\times10^{-9}$\cite{Muller}.}
\begin{ruledtabular}
\begin{tabular}{ccccc}
$\omega_i$ & $Cu_i$ & $P_{MM}$ \\
\hline
$2\omega_R +\omega_\oplus$&$[-1.13\times10^{-7}-3.01\times10^{-8}\cos\Phi_{0}$ & $-2.1(7.2)$\\
& $+8.83\times10^{-9}\sin\Phi_{0}]P_{MM}$ &\\
$2\omega_R -\omega_\oplus$&$[3.51\times10^{-8}+9.31\times10^{-9}\cos\Phi_{0}$ & $62.4(23.3)$\\
& $-2.73\times10^{-9}\sin\Phi_{0}]P_{MM}$ &\\
$2\omega_R +2\omega_\oplus$&$[4.56\times10^{-7}-1.39\times10^{-8}\cos\Phi_{0}$ & $-1.3(2.1)$\\
& $-7.08\times10^{-8}\sin\Phi_{0}]P_{MM}$ &\\
$2\omega_R -2\omega_\oplus$&$[4.37\times10^{-8}-1.34\times10^{-9}\cos\Phi_{0}$ & $-7.5(22.1)$\\
& $-6.78\times10^{-9}\sin\Phi_{0}]P_{MM}$ &
\end{tabular}
\end{ruledtabular}
\end{table}

In conclusion, we set bounds on 7 components
of the SME photon sector (Tab.\ref{Tab2}) and $P_{MM}$
(Tab.\ref{Tab3}) of the RMS framework, which are up to a factor of 7
more stringent than those obtained from previous experiments. We
have also set an upper limit [$2.1(5.7)\times10^{-14}$] on the
previously unmeasured SME component $\tilde{\kappa}_{e-}^{ZZ}$. To
further improve these results, tilt and environmental controls
will be implemented to reduce systematic effects. To remove the
assumption that $\tilde{\kappa}_{o+}$ and $\tilde{\kappa}_{e-}$
do not cancel each other, data integration will continue for more
than a year. Note added: Two other concurrent experiments have also set some similar limits \cite{Ant,Herr}.
\begin{acknowledgments}
This work was funded by the Australian Research Council.
\end{acknowledgments}

\end{document}